\documentclass[preprint,12pt]{elsarticle}

\usepackage{amssymb}
\usepackage{amsmath}
\usepackage{makecell}
\usepackage{tabularx}
\usepackage{color}
\usepackage{booktabs}
\usepackage{comment}
\usepackage{hyperref}
\usepackage{subcaption}
\usepackage{graphicx}
\usepackage{rotating}
\usepackage{array}
\usepackage{natbib}
\newcolumntype{C}{>{\centering\arraybackslash}X}

\begin{document}

\begin{frontmatter}
\begin{highlights}
\item We model the commitment-based behaviours in a two-stage optional Prisoner's Dilemma game.
\item Optionality boosts commitment acceptance but fails to sustain cooperation, causing widespread exit.
\item The strict rule promotes more cooperation by closing the ``opportunistic exit" loophole found in flexible rules.
\item The flexible rule can yield higher social welfare by allowing players to capitalise on beneficial exit options.
\item Guides the institutional design to maximise both cooperation and social welfare.
\end{highlights}

\title{Emergence of Cooperation and Commitment in Optional Prisoner's Dilemma}

\author[a]{Zhao Song}
\author[a]{The Anh Han}

\address[a]{School of Computing, Engineering and Digital Technologies, Teesside University, Middlesbrough, United Kingdom; Emails: \{Z.Song,T.Han\}tees.ac.uk}

%% Abstract
\begin{abstract}
Commitment is a well-established mechanism for fostering cooperation in human society and multi-agent systems. However, existing research has predominantly focused on the commitment that neglects the freedom of players to abstain from an interaction, limiting their applicability to many real-world scenarios where participation is often voluntary. In this paper, we present a two-stage game model to investigate the evolution of commitment-based behaviours and cooperation within the framework of the optional Prisoner's Dilemma game. In the pre-game stage, players decide whether to accept a mutual commitment. Once in the game, they choose among cooperation, defection, or exiting, depending on the formation of a pre-game commitment. We find that optional participation boosts commitment acceptance but fails to foster cooperation, leading instead to widespread exit behaviour. To address this, we then introduce and compare two institutional incentive approaches: i) a strict one (STRICT-COM) that rewards only committed players who cooperate in the game, and ii) a flexible one (FLEXIBLE-COM) that rewards any committed players who do not defect in the game. The results reveal that, while the strict approach is demonstrably better for promoting cooperation as the flexible rule creates a loophole for an opportunistic exit after committing, the flexible rule offers an efficient alternative for enhancing social welfare when such an opportunistic behaviour results in a high gain. This study highlights the limitations of relying solely on voluntary participation and commitment to resolving social dilemmas, emphasising the importance of well-designed institutional incentives to promote cooperation and social welfare effectively.  %not only encourage cooperation but also prevent its strategic avoidance.
\end{abstract}

%%Research highlights

\begin{keyword}
Commitment \sep Optional Prisoner's Dilemma game \sep Cooperation \sep Institutional incentives
\end{keyword}

\end{frontmatter}

\section{Introduction}
Commitment, in the form of contracts and promises, is a cornerstone of solving problems in human societies and multi-agent systems~\cite{akdeniz2021evolution,nesse2001evolution,frank1988passions,song2025evolution,hammond2025multi,dastani2017commitments,el2013verifying}. Empirical evidence consistently shows that establishing a prior commitment enhances the likelihood of cooperative outcomes~\cite{cherry2013enforcing,singh2014norms,johnson2006hand}. This powerful link has made commitment a focal point in the study of cooperation, a widespread evolutionary puzzle that spans numerous disciplines~\cite{dietz2003struggle,boyd2009culture}. The Prisoner's Dilemma (PD) game is the canonical framework used to investigate this puzzle, as it starkly captures the conflict between individual self-interest and collective benefit~\cite{doebeli2005models,coombs1973reparameterization,perc2013evolutionary}. In this framework, players choose whether to cooperate for a mutual gain or defect for a higher personal payoff.

While laboratory experiments consistently show that such commitments can increase cooperation rate relying on trust and repeated interactions~\cite{andersson2012credible,feldhaus2016more,balliet2010communication,dvorak2024negotiating,kapetaniou2023social}, it remains theoretically unclear given non-binding promise to cooperate does not alter the underlying incentives that make defection the rational choice~\cite{bahel2022communication,axelrod1981evolution,salahshour2019evolution}. To resolve this, a significant body of research has focused on mechanisms that make commitments credible by adding a secondary layer of enforcement~\cite{lang2024role,han2013emergence}, reputation \cite{krellner2025words}, and network reciprocity among commitment-compliant players \cite{song2025network,flores2024evolution}. These studies demonstrate that through institutional or peer-based incentives---such as rewarding compliance or punishing violations---or by fostering associations among promise-keeping players, individuals find it profitable to honour their cooperative commitments, thus establishing a strong pathway to cooperation in one-shot social dilemma interactions~\cite{han2016emergence,duong2021cost,sasaki2015commitment,han2016synergy}. 

However, existing research on commitment has predominantly been conducted under the assumption of compulsory participation, where players are captive to the interaction~\cite{han2022institutional,locey2012commitment}. This overlooks a critical feature of many real-world scenarios: the freedom to abstain~\cite{shen2021exit,lin1990collectivization,nosenzo2017effect}. In the various scenarios, including business, social life, and international relations, individuals or groups can often freely choose to exit an interaction, possibly after making a joint commitment. Voluntary participation is demonstrated to fundamentally reshape the strategic dynamics but has been largely neglected in the context of commitment mechanisms~\cite{sharma2023small,yamamoto2019effect,szabo2002evolutionary}. Arguably, the ability to exit the game without consequence after agreeing to a joint commitment may encourage players to accept it, thereby increasing participation levels in the commitment process---an aspect that has been shown to be critical for realising the benefits of commitment-based mechanisms \cite{han2022institutional}. Therefore, it is crucial to understand how commitment and the supporting institutional mechanisms operate when faced with the freedom to exit.

In this paper, we address the gap by developing and analysing a two-stage evolutionary game model that integrates (prior) commitment with the optional Prisoner's Dilemma (OPD). We first show that the option to exit creates a paradox: it significantly boosts players' willingness to accept a commitment but fails to foster subsequent cooperation, leading instead to widespread abandonment of the game. To resolve this, we introduce and compare two approaches regarding how institutional incentive is provided for commitment-compliant players: STRICT-COM, a strict rule that rewards only committed cooperation, and FLEXIBLE-COM, a flexible rule that rewards any committed, non-defecting action. 
Our results reveal a nuanced trade-off. The strict rule is more effective at promoting cooperation because it closes the ``opportunistic exit" loophole where players commit only to claim a reward, which, conversely, makes the flexible rule vulnerable. However, the rigidity of the strict rule can be a disadvantage, as it may prevent players from capitalising on a highly beneficial outside option. In contrast, the flexible rule, by tolerating exit, can lead to higher social welfare precisely in those scenarios where exit is beneficial and the incentive budget is limited. This study thus provides critical insights for designing institutions in voluntary settings, demonstrating that the optimal mechanism depends on the overarching goal of both cooperation and social welfare.

\section{Models and Method}
We first recall the optional Prisoner’s Dilemma (OPD) game, based on which we then describe our model that integrates prior commitments and institutional incentives. 

\subsection{Optional Prisoner's Dilemma (OPD)}

In the OPD, players have three possible actions: cooperation ($C$), defection ($D$), and exit ($L$). If both players cooperate, each receives the same reward $R$. If both players defect, each receives the same penalty $P$. In the case where one player cooperates while the other defects, the former player receives a sucker's payoff $S$, and the latter player receives the temptation to defect $T$. Finally, if at least one player chooses to exit the game, the interaction concludes without any exchange, resulting in both players receiving a payoff $\sigma$.  
Thus, OPD can be described with the following payoff matrix:
\begin{equation}
\begin{array}{c@{\quad}|ccc}
& C & D & L \\
\hline
C & R & S & \sigma \\
D & T & P & \sigma\\
L & \sigma & \sigma & \sigma 
\end{array}.
\end{equation}
In line with previous work, the exit payoff is set so that it is higher than the penalty from mutual defection, but lower than the reward of mutual cooperation, namely, $T>R>\sigma >P>S$ \citep{janssen2008evolution}. 
For a clear comparison with previous work while focusing on the impact of voluntary participation (on commitment-based behaviours), we fix $R = 1$, $S=-1$, $T = 2$, and $P = 0$~\cite{han2022institutional} and vary $\sigma$  within the interval $[0,1]$. 

In the OPD, cooperation is maintained by the cyclic dynamic among three strategies, where cooperation is invaded by defection, exit is advantageous to defection, followed by the replacement by cooperation \cite{hauert2002volunteering}. 

\begin{figure}[htb]
     \centering
\includegraphics[width=\linewidth]{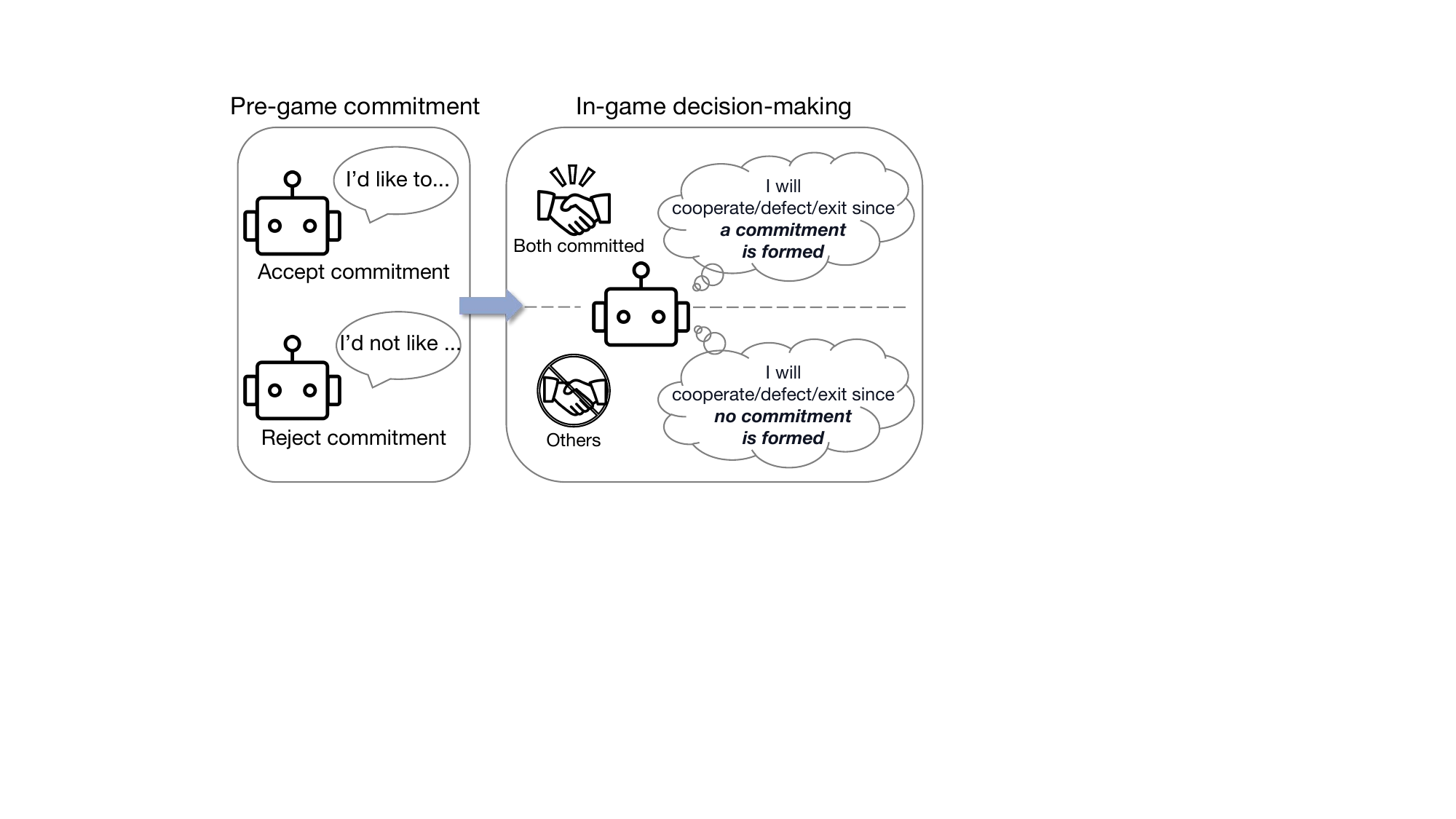}
    %\vspace{0.1cm}
    \caption{
    \textbf{Two-stage game involving commitment within the framework of the optional Prisoner's Dilemma (OPD).} 
    For the pre-game commitment stage, players choose whether to accept to join a commitment ($A$) or not ($N$). The commitment is formed only when both players accept. For the in-game decision-making stage, players have three options: cooperation ($C$), defection ($D$), and exit ($L$). They can choose which action to take conditionally on whether the commitment was formed in the pre-game stage, thus resulting in a total of 18 possible strategies in this two-stage game.}
    %\vspace{0.2cm}
    \label{fig: model}
\end{figure}

\subsection{Optional Prisoner's Dilemma with commitment}

In this study, we propose a novel model that integrates prior commitments and OPD in a two-stage non-repeated interaction, see Figure \ref{fig: model} for an illustration. Before the game, each player independently decides whether to accept to join a commitment ($A$) or not ($N$). If both players choose to accept, a (non-binding) commitment is formed. In that case, they incur a cost of $\varepsilon$, representing the effort to maintain the commitment. However, if at least one player opts not to commit, no commitment is formed. Then players engage in the OPD, where they can adopt one of the three actions, cooperation ($C$), defection ($D$), or exit ($L$), depending on whether the commitment is formed (that is, they might choose the same or different actions when the commitment is formed vs not formed). Each player's strategy can be represented using the notation $XYZ$, where $X$ indicates whether the player accepts to join a commitment or not, taking the values of either $A$ or $N$; The variables $Y$ and $Z$ represent the actions chosen when a commitment is formed or not, taking the values of $C$, $D$, or $N$, respectively. In total, we define eighteen possible strategies derived from the combinations of their pre-game commitments and in-game strategies (Table \ref{tab:18strategy}). For example, $ACD$ accepts to join a commitment, cooperates in the presence of a formed commitment, and defects in its absence. For  the complete payoff matrix, see Table \ref{PayoffMatrix} in Appendix.
\begin{table*}[htb]
    \centering
    \caption{Eighteen strategies in the OPD with prior commitment formation.}
\includegraphics[width=\linewidth]{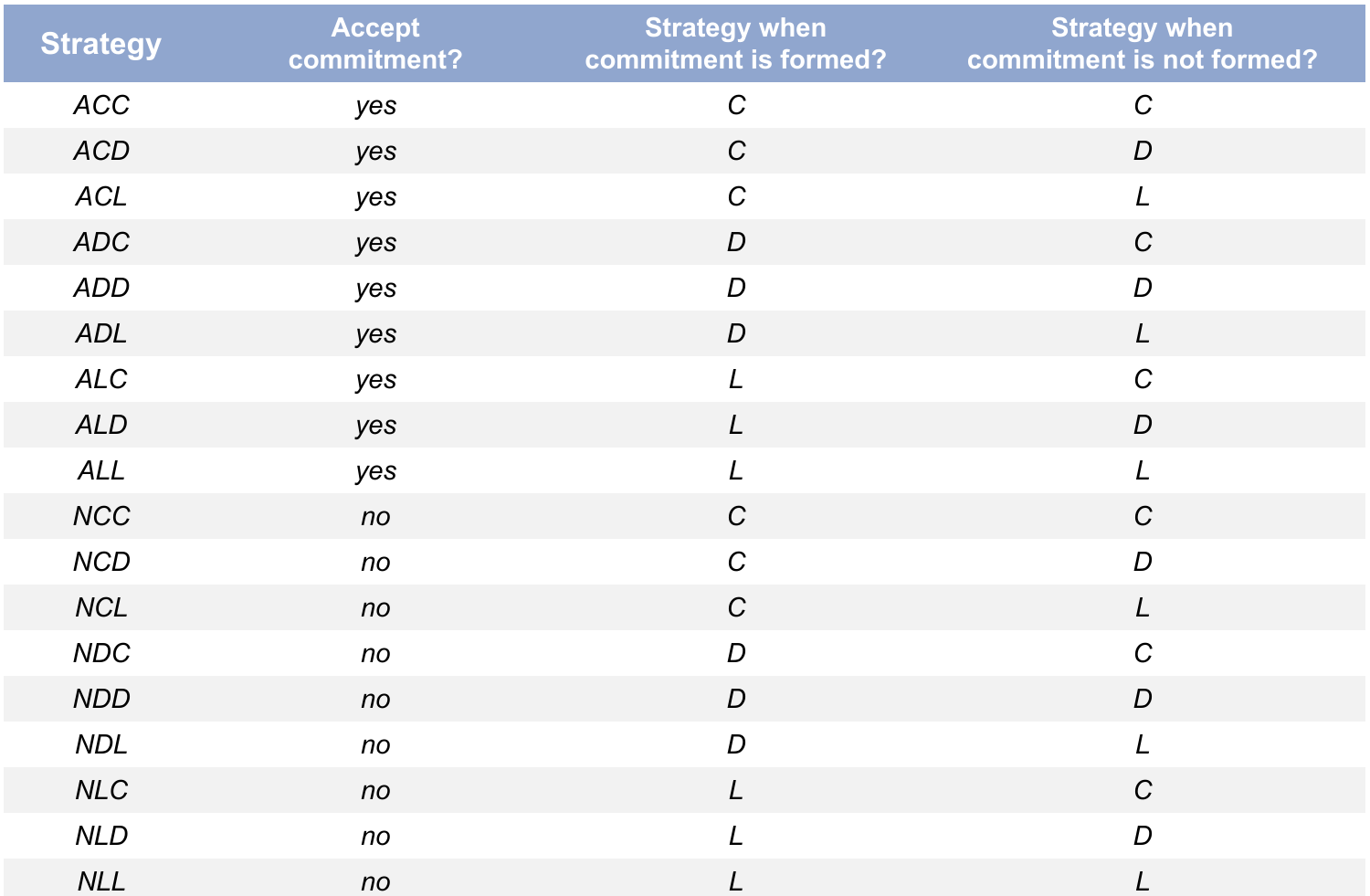}
   \label{tab:18strategy}
\end{table*}

% \begin{table*}[htb]
%     \centering
%     \caption{Eighteen strategies in the OPD with prior commitment formation.}
%     \begin{tabular}{c|ccc}
%     \hline
%      Strategy   & \makecell{Accept \\ commitment?} & \makecell{Strategy when commitment \\is formed?} & \makecell{Strategy when commitment \\ is not formed?}\\
%      \hline
%     $ACC$     & yes & $C$ & $C$ \\
%     $ACD$     & yes & $C$ & $D$ \\
%     $ACL$     & yes & $C$ & $L$ \\
%     $ADC$     & yes & $D$ & $C$ \\
%     $ADD$     & yes & $D$ & $D$ \\
%     $ADL$     & yes & $D$ & $L$ \\
%     $ALC$     & yes & $L$ & $C$ \\
%     $ALD$     & yes & $L$ & $D$ \\
%     $ALL$     & yes & $L$ & $L$ \\
%     $NCC$     & no & $C$ & $C$ \\
%     $NCD$     & no & $C$ & $D$ \\
%     $NCL$     & no & $C$ & $L$ \\
%     $NDC$     & no & $D$ & $C$ \\
%     $NDD$     & no & $D$ & $D$ \\
%     $NDL$     & no & $D$ & $L$ \\
%     $NLC$     & no & $L$ & $C$ \\
%     $NLD$     & no & $L$ & $D$ \\
%     $NLL$     & no & $L$ & $L$ \\
%     \hline
%     \end{tabular}
%     \label{tab:18strategy}
% \end{table*}

\subsection{Institutional incentives}
We assume that there is a \textit{per capita} budget $u$ available for providing incentives, rewarding commitment-compliant players (whenever a commitment is formed). Extended from the assumption mainly focuses on the commitment to cooperate ~\cite{han2022institutional}, here two types of commitment are considered: i) STRICT-COM, which requires committed players to cooperate, and ii) FLEXIBLE-COM, which requires committed players not to defect. The latter is more flexible than the former, allowing committed players to  exit the game without violating the commitment terms and conditions.
Accordingly, institutional reward is implemented depending on whether the in-game strategy is consistent with their pre-game commitment:
\begin{itemize}
    \item STRICT-COM: Commit to cooperate and cooperate when commitment is formed, receiving a reward $u$ with $X=A$ and $Y=C$, including $ACC$, $ACD$, and $ACL$, see payoff matrix in Table  \ref{table:incentive1}.
    \item FLEXIBLE-COM: Commit not to defect, and cooperate or exit when a commitment is formed, receiving a reward $u$ with $X=A$ and $Y\neq D$, including $ACC$, $ACD$, $ACL$, $ALC$, $ALD$, and $ALL$, see payoff matrix in  Table \ref{table:incentive2}.
\end{itemize}

\subsection{Evolutionary dynamics}
We consider the well-mixed finite population of $M$ players, where the evolutionary dynamics is shaped by social learning ~\cite{nowak2004emergence}.
Each player updates their strategy following the Moran process. At each time step, a randomly selected player updates its strategy by imitating the strategy of another randomly selected player. 

In particular, for a simplified case with two strategies, we assume $m_i$ players adopt strategy $i$ and the other $M-m_i$ players adopt strategy $j$ in the population, which can be one of the eighteen strategies (see  Table \ref{tab:18strategy}). Then the average payoffs for $i$ and $j$ are:
\begin{equation}
\begin{split}
     & P_{i,j}=\frac{(m_i-1)\pi_{i,i}+(M-m_i)\pi_{i,j}}{M-1}, \\
     & P_{j,i}=\frac{m_i\pi_{j,i}+(M-m_i-1)\pi_{j,j}}{M-1},
\end{split}
\end{equation}
where $\pi_{i,j}$ is the payoff when $i$ interacts with $j$.
In a finite population, such average payoffs represent individual fitness, or their social success.
We assume the evolutionary dynamic is driven by the Fermi function \cite{blume1993statistical}, one of the typical social learning rules where players tend to imitate the strategy of the more successful player more often~\cite{sigmund2010calculus}. In particular, the player with payoff $P_{i,j
}$ will  adopt the strategy of the pairwise player with payoff $P_{j,i}$ with a probability: $(1+e^{s (P_{i,j}-P_{j,i})})^{-1}$. Herein, $s$ represents the intensity of selection, determining how strongly players rely on the fitness difference when making their imitation decision. When $s\rightarrow \infty$, imitation is deterministic, while when $s=0$, it becomes the neutral draft where players randomly adopt strategies from pairwise players.

Therefore, the probability of the number $m$ of strategy $i$ in the population increasing or decreasing by 1 is:
\begin{equation}
       T_{i,j}^{\pm}=\frac{M-m_i}{M}\frac{m_i}{M}[1+e^{\mp s (P_{i,j}-P_{j,i})}]^{-1}.
\end{equation}
The fixation probability of a mutant with strategy $i$ in a resident population of $M-1$ $j$-players is \cite{traulsen2006stochastic}:
\begin{equation}
    \rho_{j,i}=\frac{1}{1+\sum_{k=0}^{M-1}\prod_{m_i=1}^{k}\frac{T^-_{i,j}}{T^+_{i,j}}}.
\end{equation}
Assuming a small mutation limit, where any mutant either fixates or goes extinct before another mutation occurs \cite{nowak2004emergence,imhof2005evolutionary},   the fixation probabilities $\rho_{ij}$ define the transition probabilities of the Markov process between 18 different homogeneous states of the population. It has demonstrated a broad range of applicability that extends well beyond the narrow threshold of very small mutation (or exploration) rates \cite{hauert2007via,rand2013evolution,sigmund2010social,han2013emergence}.

The transition matrix with $T_{i,j(i\neq j)}=\rho_{i,j}/(q-1)$ and $T_{ii}=1-\sum_{j=1, j\neq i}^qT_{ij}$, where $q$ is the number of strategy. The normalised eigenvector associated with the eigenvalue 1 of the transition matrix provides the stationary distribution, describing the relative time the population spends adopting each of the strategies.
% \paragraph{Risk-dominance}
% Risk dominance is a powerful tool to discriminate the transition direction between two strategies $i$ and $j$ ~\citep{sigmund2010calculus}. Comparing a mutation $i$ in the fixation population using strategy $j$, and a mutation $j$ in the fixation population using strategy $i$, the former is the risk dominant strategy when, 
% \begin{equation}
%     \pi_{i,i}+\pi_{i,j}>\pi_{j,j}+\pi_{j,i}.
%     \label{risk}
% \end{equation}

\subsection{Social Welfare}
Social welfare \citep{han2024evolutionary}, the total population payoff taking into account the cost of providing incentives, is given as follows
\begin{equation}
    SW=\sum_{n\in S} \pi_{n,n}p_n,
\end{equation}
where $n$ is the state of population, $S$ is the all possible states (the collection of strategies), $\pi_{n,n}$ is the average payoff when the population is at state $n$, and $p_n$ is the probability that the population is at state $n$. Note that in our calculation, we assume that the rewarded player's benefit is the same as the cost incurred by the rewarding institution. Thus,   rewards and institutional costs can be excluded from the calculation of social welfare.

\begin{figure*}[t!]
\centering
\includegraphics[width=\linewidth]{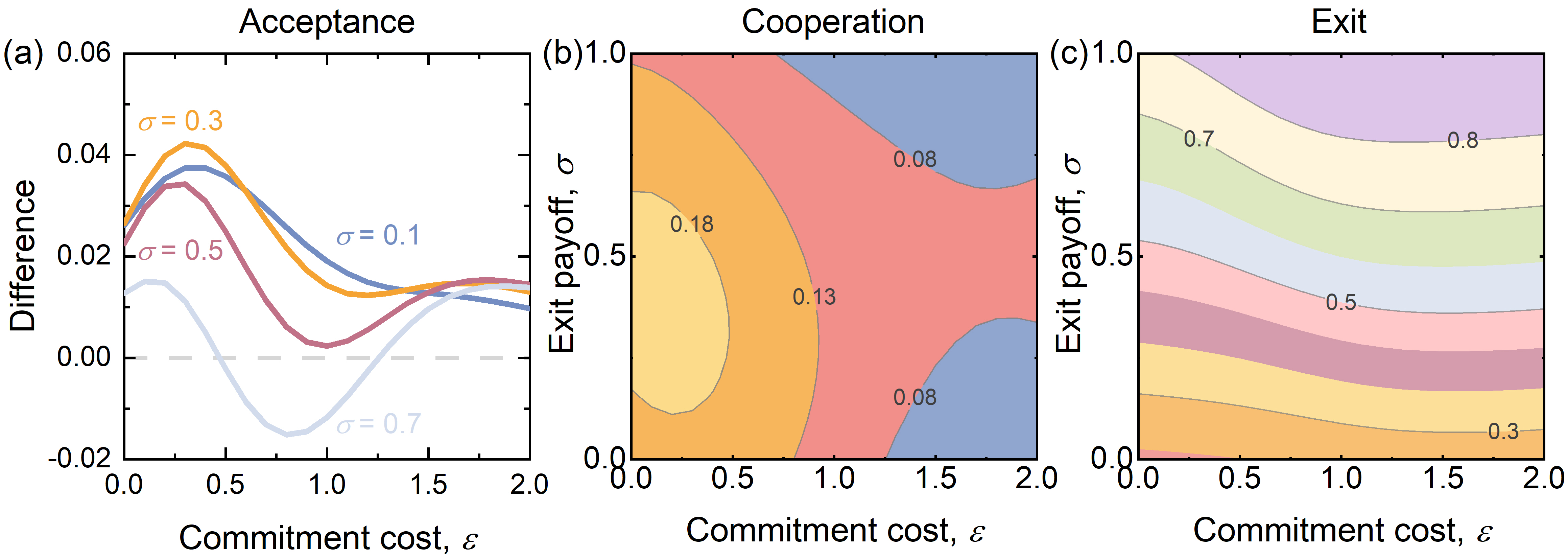}
    \caption{
    \textbf{While the option to exit enhances commitment acceptance compared to the PD, it fails to sustain cooperation and leads to prevalent exit.}
    Panel (a) shows the difference in commitment acceptance rate between OPD and PD. Panel (b) shows the frequency of overall cooperation, and panel (c) shows the frequency of overall exit in OPD. Both (b) and (c) are plotted as a function of commitment cost $\epsilon$ and exit payoff $\sigma$, respectively.
    Other parameters: $s=0.1$, $M = 100$. \\
    }
    % \vspace{0.2cm}
    \label{fig1}
\end{figure*}

\section{Results}
\subsection{The paradox of optional participation: commitment without cooperation}
In one-shot OPD scenarios, while the ability to exit without consequence is highly effective at encouraging players to commit, it simultaneously undermines the potential for cooperation, leading to widespread exit from the game. Specifically, the OPD proves more effective than the classical PD framework at securing an initial commitment. As shown in Figure \ref{fig1}(a), OPD maintains superior commitment acceptance across a wide range of commitment costs, an advantage that only vanishes when the exit payoff becomes so high that participating in the game itself is an unattractive proposition. Though the difference is slight, the enhancement is significant given the low baseline acceptance rate in the PD \cite{han2022institutional} and especially strong for high commitment costs (see Appendix, Figure \ref{figA1}). 

Nevertheless, this high rate of commitment acceptance does not translate into subsequent cooperation. The overall cooperation rate is critically low, diminishing to near zero as soon as a minimal commitment cost is introduced, as shown in Figure \ref{fig1}(b). Instead of fostering cooperation, the primary effect of OPD is to enable exit as the dominant strategy. As shown in Figure \ref{fig1}(c), the decision to leave the game is governed almost exclusively by the exit payoff; its frequency rises as its payoff increases. With less effectiveness, a higher commitment cost also induces more exit.
This indicates that the exit option makes players likely to commit, but they will preferentially choose the certainty of a guaranteed exit payoff over the risks and costs inherent in the PD social dilemma.

\begin{figure*}[h!]
\centering
\includegraphics[width=\linewidth]{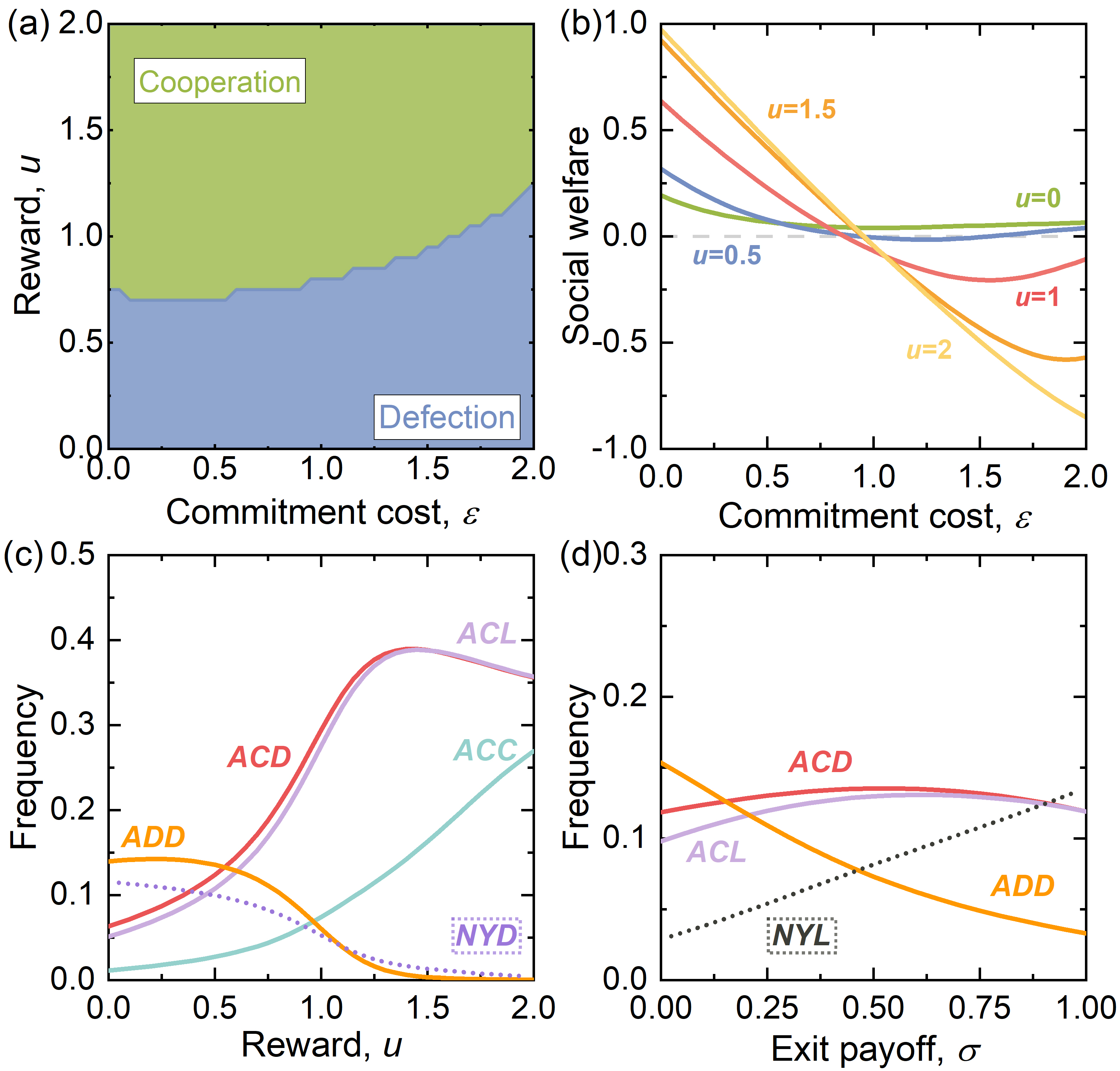}
    \caption{
    \textbf{
    The STRICT-COM incentive fosters cooperation and social welfare by establishing conditional cooperation as the dominant evolutionary strategy.}
    Panel (a) shows the region where cooperation, defection, and exit dominate in the population, respectively. 
    Panel (b) shows the social welfare as a function of commitment cost $\varepsilon$ for various rewards $u$.
    Panels (c) and (d) show the frequency of prevalent strategies (those with a frequency $>0.1$) as a function of reward and exit payoff, respectively. 
    Parameters are set as $s=0.1$; (a-b): $\sigma=0.1$; (c): $\varepsilon=0.1$ and $\sigma=0.1$; (d): $u=0.5$ and $\varepsilon=0.1$.\\}
    \label{fig2}
\end{figure*}

\subsection{Institutional incentives enhance cooperation as well as social welfare}

To address the limitation of commitment in sustaining cooperation in the OPD, we next examine the two proposed approaches for providing institutional incentives. As shown in the following, while both incentives can successfully promote cooperation, as well as social welfare, their specific design dictates the robustness against strategic exploitation.

STRICT-COM is highly effective at fostering cooperation and enhancing social welfare by making committed conditional cooperation the most successful strategy. This success is driven by institutional reward, acting as a critical switch for population behaviour. As shown in Figure \ref{fig2}(a), a moderate reward (around $u>0.75$) shifts the evolution from a state dominated by defection to one with widespread cooperation, a transition that holds even when the commitment cost is high. Such behavioural shifts directly improve social welfare, which is maximised when rewards are high and commitment costs are low (left-top corner in Figure \ref{fig2}(b)). However, these benefits are fragile; high commitment costs erode social welfare, eventually causing a net loss even in the cooperative phase (right-top corner in Figure \ref{fig2}(b)).

From the perspective of strategy evolution, the success of the STRICT-COM is rooted in its ability to reshape the evolutionary landscape, establishing conditional cooperation as the dominant behaviour while marginalising unconditional defection. As the reward increases, the frequency of fake defectors ($ADD$, which pretends to commit but always defects) collapses, while conditional cooperators ($ACD$ and $ACL$) rise to become the dominant strategies (Figure \ref{fig2}(c)). These strategies, which anchor their cooperation to the presence of commitment, form the backbone of the cooperative state. Meanwhile, the frequency of committing unconditional cooperators ($ACC$, which always accept commitment and cooperate) increases as the reward rises.
Furthermore, the availability of an outside option further reinforces this pro-social shift. As the exit payoff increases, it suppresses the fake defective strategy $ADD$, while the dominant conditional cooperators maintain their high frequency (Figure \ref{fig2}(d)). Our additional analyses show that these remarkable observations are robust for other parameter values of $\sigma$ and $\epsilon$, see Figures \ref{figA2} and \ref{figA4} in Appendix.

\begin{figure*}[h!]
\centering
\includegraphics[width=\linewidth]{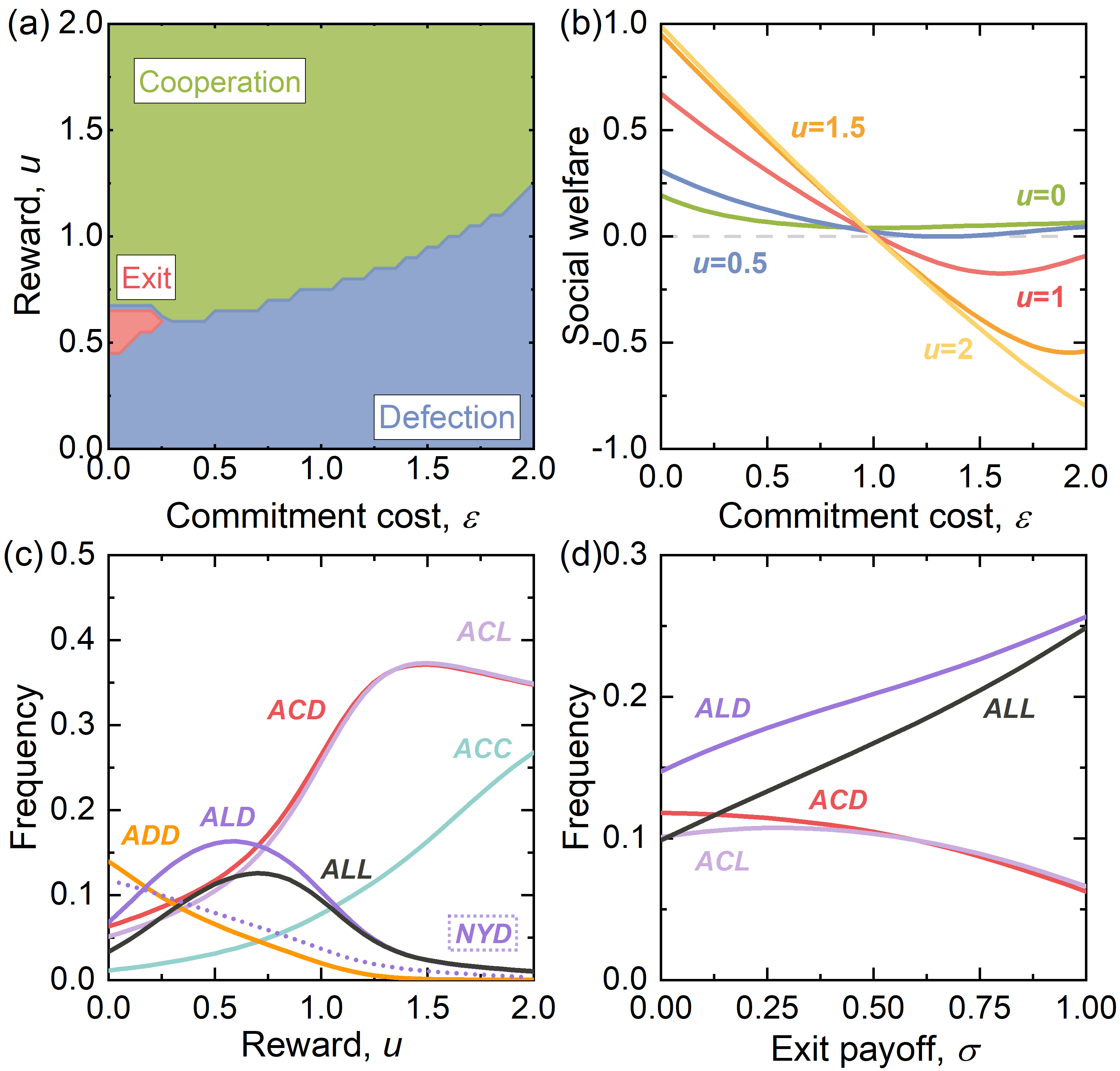}
    \caption{
    \textbf{
    The FLEXIBLE-COM incentive fosters cooperation and social welfare, but its flexible nature enables exit to emerge as a dominant strategy.}
    Panel (a) shows the region where cooperation, defection, and exit dominate in the population, respectively. 
    Panel (b) shows the social welfare as a function of commitment cost $\varepsilon$ for various rewards $u$.
    Panels (c) and (d) show the frequency of prevalent strategies (those with a frequency $>0.1$) as a function of reward and exit payoff, respectively.
    Parameters are set as $s=0.1$; (a-b): $\sigma=0.1$; (c): $\varepsilon=0.1$ and $\sigma=0.1$; (d): $u=0.5$ and $\varepsilon=0.1$.\\}
    \label{fig3}
\end{figure*}

Now, FLEXIBLE-COM also promotes cooperation and enhances social welfare, but its flexible nature introduces a critical vulnerability: it encourages players to strategically exit, making the system less robust than its strict counterpart (in terms of sustaining cooperative behaviour). Similar to the strict rule, a moderate institutional reward induces a shift from defection to cooperation (Figure \ref{fig3}(a)). However, a key difference emerges at a low reward and moderate commitment cost, where a new phase appears, dominated by exit. This creates a more complex strategic environment, and while social welfare is similarly maximised when rewards are high and costs are low, the benefits are once again severely eroded by increasing commitment costs (Figure \ref{fig3}(b)). 

This outcome is a direct result of the strategic exit. Under the flexible rule, which rewards any non-defecting action in the presence of commitment, the new opportunistic strategies ($ALL$ and $ALD$, which exit in the presence of commitment) become highly viable as they seize the increased exit benefit without consequence. As shown in \ref{fig3}(c), conditional cooperators ($ACL$ and $ACD$) still rise in frequency with increasing reward; they are now in direct competition with these exit strategies, where players commit and simply exit, claiming a reward without participating in the game. The most dramatic divergence from the strict mechanism is revealed by the role of the exit payoff. Instead of suppressing defection, a higher exit payoff now actively promotes the rise of exit-based strategies like $ALL$ and $ALD$ (Figure  \ref{fig3}(d)). This comes at the direct expense of conditional cooperation, as the frequencies of $ACD$ and $ACL$ now decline with a higher exit payoff. This reveals the double-edged nature of the flexible rule: by tolerating exit, it transforms the outside option from an undesirable alternative into a dominant strategic pathway, fundamentally altering the dynamics of cooperation. Results for various values of exit payoff are shown in Figures \ref{figA3} and \ref{figA5}.

\begin{figure*}[tb]
\centering
\includegraphics[width=\linewidth]{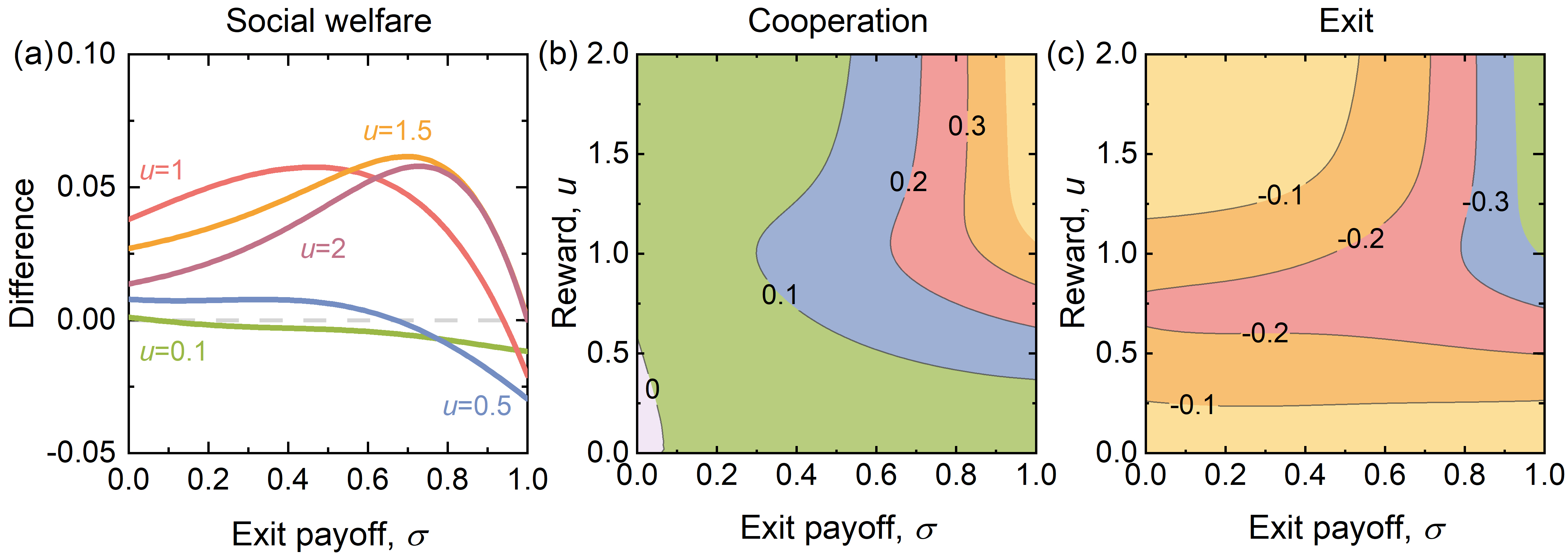}
    \caption{
    \textbf{Comparatively, STRICT-COM is more effective for promoting cooperation, whereas FLEXIBLE-COM can achieve higher social welfare when the exit benefit ($\sigma$) is high and the incentive budget  ($u$) is limited.}
    Panel (a) shows the difference in social welfare between STRICT-COM and FLEXIBLE-COM as a function of exit payoff. Panels (b) and (c) show the difference in cooperation and exit between STRICT-COM and FLEXIBLE-COM as a function of exit payoff and reward, respectively. 
    Parameters are set as $\varepsilon=0.1$ and $s=0.1$.}
    % \vspace
    \label{fig4}
\end{figure*}

%\subsection{STRICT-COM proves superior to FLEXIBLE-COM by preventing strategic exit.}
\subsection{Strictness versus Flexibility }
The direct comparison of the two institutional incentive approaches confirms that STRICT-COM is generally more effective and robust than its FLEXIBLE counterpart for promoting cooperation. 
As shown in Figure \ref{fig4} (b), this advantage is evident in the consistently higher frequency of cooperative behaviours observed under the strict rule, particularly when the reward is high and the exit option is moderately attractive. Furthermore, the frequency of exit is substantially lower under the strict rule. This highlights the core design difference: by exclusively rewarding cooperation, STRICT-COM closes the ``opportunistic exit" loophole—a pathway where players commit but exit, claiming an institutional reward without ever risking defection—that FLEXIBLE-COM inadvertently creates. This makes the strict rule a more robust mechanism for ensuring that commitment translates into pro-social action.

However, when considering the population's social welfare instead of its cooperation level as the optimisation objective---as these two objectives are often misaligned \cite{han2024evolutionary}---FLEXIBLE-COM can be superior, especially when the exit benefit is high and the incentive budget is low. As shown in Figure 4(a), when the institutional reward is high, STRICT-COM typically yields greater social welfare, particularly when the exit payoff is low to moderate. This advantage, however, diminishes and ultimately reverses as the exit payoff grows, with FLEXIBLE-COM becoming the more beneficial strategy when leaving the game is highly profitable. Moreover, when the reward budget is limited ($u\leq0.5$), the advantage shifts decidedly to FLEXIBLE-COM, which outperforms the strict rule across a much broader range of exit payoffs. This reversal occurs because the strict rule's primary strength—closing the ``opportunistic exit" loophole—becomes a liability when the exit option is highly profitable. By preventing players from benefiting from exiting, especially with limited reward budgets, STRICT-COM caps the potential social welfare. In contrast, the flexible rule, by tolerating this very loophole, allows the population to achieve a greater social welfare precisely when exiting is the most profitable action. 

It is crucial to note, however, that this trade-off is itself sensitive to the cost of commitment. As shown in Figure \ref{fig4-3}, when this cost increases, the advantage shifts decisively towards the STRICT-COM rule, which then promotes higher social welfare across a broader range of conditions. This result further reinforces our main conclusion: there is no one-size-fits-all solution. Optimal institutional design requires a careful, context-dependent approach, balancing the goals of enhancing cooperation and welfare by considering the interplay between the incentive budget, the value of outside opportunities, and the very costs of participating in the commitment.

\section{Discussion}
We propose a two-stage game model to investigate the evolution of commitment and cooperation within the context of the optional Prisoner's Dilemma game. We find that while the freedom to exit encourages players to accept to join a commitment more often---which is key to ensure the benefit of commitment-based mechanisms \cite{nesse2001evolution,han2022institutional}---it paradoxically fails to foster cooperation, leading instead to a population that overwhelmingly chooses to opt out of the interaction in one-shot scenarios. To address this deficiency, we explored two approaches to institutional incentives: STRICT-COM and FLEXIBLE-COM, which effectively sustain cooperation. Interestingly, our direct comparison between these incentive approaches reveals that though beneficial, the latter is vulnerable to an ``opportunistic exit" whenever the exit benefit is sufficiently high,  due to the flexible nature of the approach. The strict rule proves superior for enabling cooperation precisely due to the inhibition of this loophole by exclusively rewarding pro-social action, resulting in higher levels of both cooperation and social welfare. However, when focusing on population social welfare as the optimisation objective \cite{han2024evolutionary}, FLEXIBLE-COM provides an efficient alternative to STRICT-COM  whenever the exit benefit is high and the incentive budget is low. 
Thus, our analysis suggests that when designing institutional incentives for maximising the population/system social welfare, commitments need to be flexible to account for players' ``opportunistic exit" behaviour to reap the benefits of the alternative choices.
This principle is especially relevant when the game environment might change over time, where new alternatives might emerge and the game payoff structure might change after a commitment was formed \cite{teague2002investigating,el2013verifying}. 

We demonstrate that within an optional participation framework, a high willingness to commit does not inherently translate into cooperative action, necessitating the use of institutional incentives. Our finding that players readily accept commitments but subsequently fail to cooperate aligns with long-standing experimental evidence showing that promises made before the game are often not honoured ~\cite{chen1994effects}. Furthermore, our work reinforces the conclusion from previous studies that institutional or peer-based incentives are essential for ensuring compliance with cooperative agreements, moving beyond mere acceptance to achieve pro-social outcomes~\cite{han2022institutional,sasaki2015commitment}. Our contribution is to show that this principle holds even in the volatile context of optional participation, where the institution must compete not only against defection but also against the choice to exit the game entirely.

Furthermore, our results reveal a clear misalignment between the goals of maximising cooperation and social welfare. We found that under both the STRICT-COM and FLEXIBLE-COM incentives, it is possible to create conditions where cooperation is the dominant strategy, yet high commitment costs lead to a negative net payoff for the population. This provides a concrete example supporting the broader principle that evolutionary mechanisms designed to promote cooperation may not necessarily improve collective well-being ~\cite{han2024evolutionary}. By comparing the two incentive structures, our work takes this insight a step further, demonstrating how specific institutional designs can manage this trade-off. We show that the superior performance of the strict rule is not just in its ability to foster cooperation, but in its efficiency at doing so without creating costly loopholes like the ``opportunistic exit" seen in the flexible rule.

Our analysis provides new understanding of commitment in optional interactions, but its setting in a well-mixed, finite population naturally opens several avenues for future research. First, extending this model to structured populations is a crucial next step. In such settings, where interactions are local, cooperation can be sustained through network reciprocity \cite{perc2013evolutionary}, and the dynamics of commitment and cooperation could be fundamentally different. Second, our model could be adapted to study hybrid populations of human agents and pre-programmed bots. Recent work has shown that even simple bots (e.g., ``loner bots" who always exit) can significantly alter the evolutionary landscape in optional games~\cite{sharma2023small}. Investigating how a small fraction of such principled bots shapes commitment and cooperation could reveal powerful new ways to solve social dilemmas. Finally, a promising direction is the design of adaptive institutions. Rather than using a fixed rule, an institution could dynamically switch between punishment and reward, using a low-cost ``carrot" when cooperation is high and a more targeted ``stick" when it falters~\cite{andreoni2003carrot,chen2015first}, could optimise the alignment between cooperation and social welfare in real-time.

\section*{Acknowledgments}
We acknowledge the support provided by EPSRC (grant EP/Y00857X/1) to Z.S. and T.A.H.

\bibliographystyle{elsarticle-num} 
\bibliography{mybib}

\newpage
\section{Appendix A}
\setcounter{figure}{0}
\renewcommand{\thefigure}{A\arabic{figure}}

\begin{figure}[htb]
\centering
\includegraphics[width=0.6\linewidth]{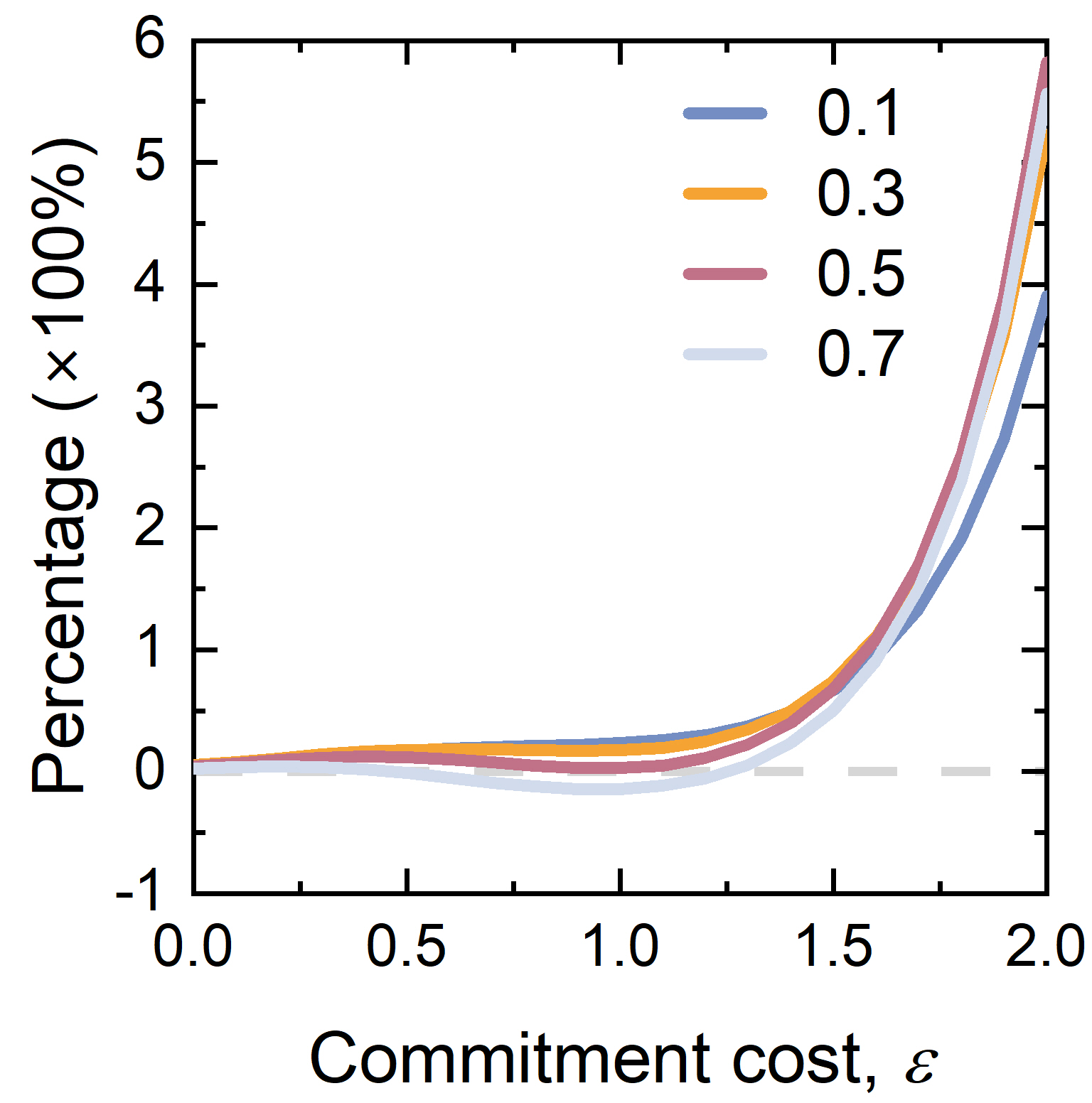}
    \caption{
    \textbf{
    Improvement percentage.}
    The parameter is set as $s=0.1$.\\}
    \label{figA1}
\end{figure}

\begin{figure}[htb]
\centering
\includegraphics[width=\linewidth]{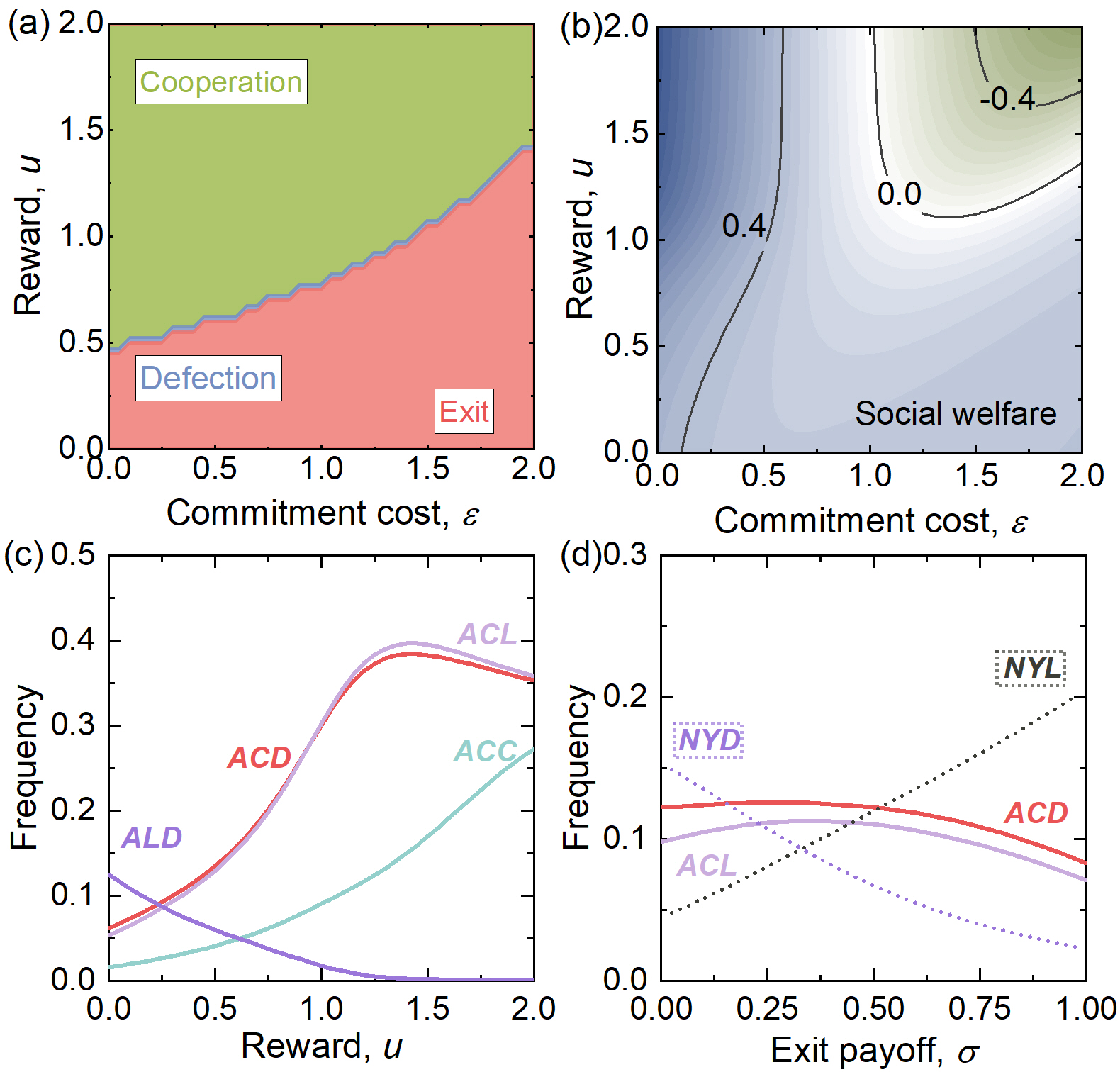}
    \caption{
    \textbf{
    STRICT-COM mechanism.}
    Panel (a) shows the region where cooperation, defection, and exit dominate in the population, respectively. 
    Panel (b) shows the social welfare as a function of commitment cost $\varepsilon$ and reward $u$.
    Panels (c) and (d) show the frequency of strategy as a function of reward and exit payoff, respectively.
    % The dark lines represent the results of OPD with commitment, while the light lines represent the results of the PD with commitment.
    % as shown in Fig. \ref{fig:baseline2}. 
    Parameters are set as $s=0.1$; (a-b): $\sigma=0.5$; (c): $\varepsilon=0.1$ and $\sigma=0.5$; (d): $u=0.5$ and $\varepsilon=0.5$.\\}
    \label{figA2}
\end{figure}

\begin{figure}[htb]
\centering
\includegraphics[width=\linewidth]{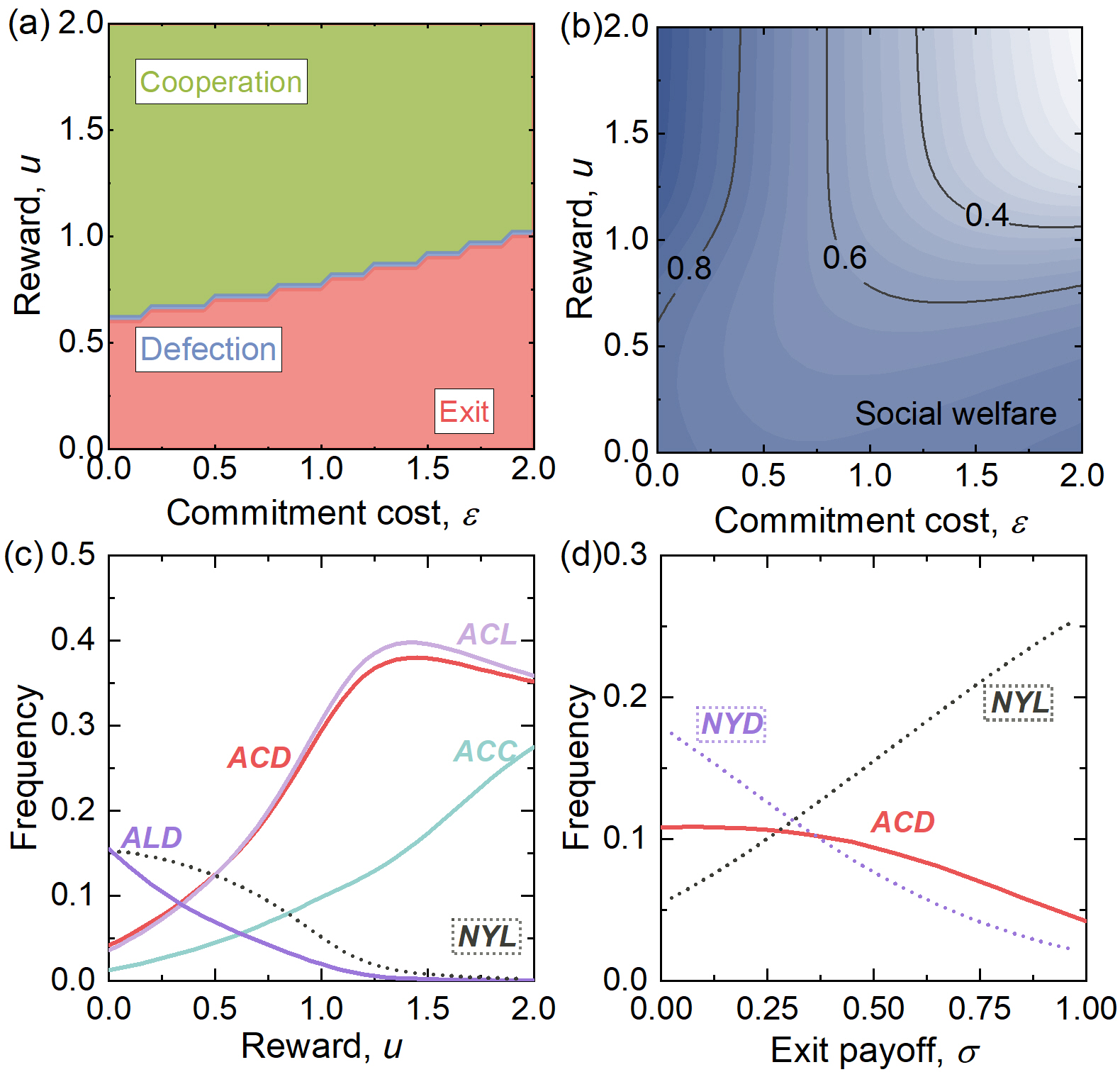}
    \caption{
    \textbf{
    STRICT-COM mechanism.}
    Panel (a) shows the region where cooperation, defection, and exit dominate in the population, respectively. 
    Panel (b) shows the social welfare as a function of commitment cost $\varepsilon$ and reward $u$.
    Panels (c) and (d) show the frequency of prevalent strategies (those with a frequency $>0.1$) as a function of reward and exit payoff, respectively.
    % The dark lines represent the results of OPD with commitment, while the light lines represent the results of the PD with commitment.
    % as shown in Fig. \ref{fig:baseline2}. 
    Parameters are set as $s=0.1$; (a-b): $\sigma=0.9$; (c): $\varepsilon=0.1$ and $\sigma=0.9$; (d): $u=0.5$ and $\varepsilon=0.9$.\\}
    \label{figA4}
\end{figure}

\begin{figure}[htb]
\centering
\includegraphics[width=\linewidth]{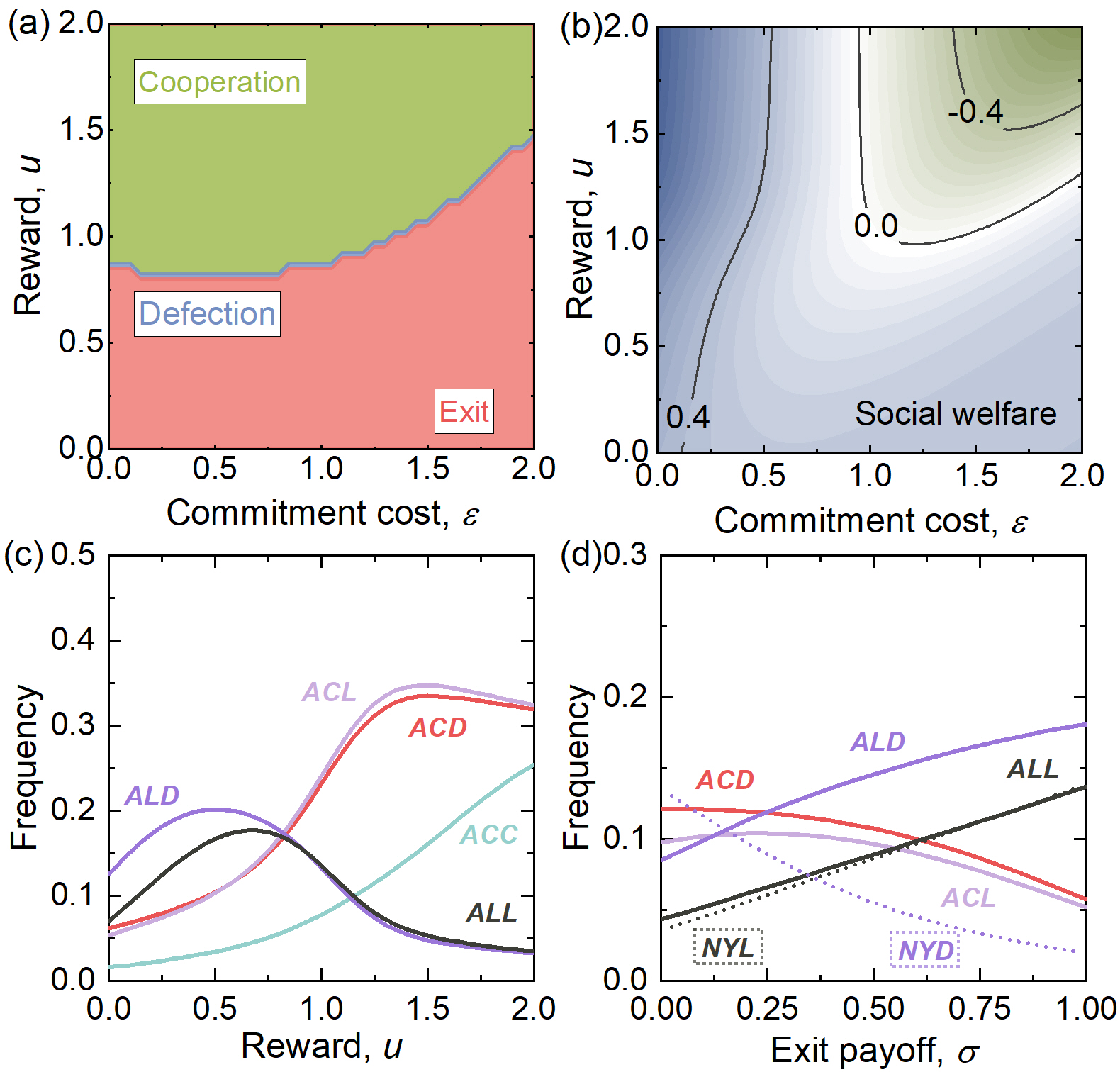}
    \caption{
    \textbf{
    FLEXIBLE-COM mechanism.}
    Panel (a) shows the region where cooperation, defection, and exit dominate in the population, respectively. 
    Panel (b) shows the social welfare as a function of commitment cost $\varepsilon$ and reward $u$.
    Panels (c) and (d) show the frequency of prevalent strategies (those with a frequency $>0.1$) as a function of reward and exit payoff, respectively.
    % The dark lines represent the results of OPD with commitment, while the light lines represent the results of the PD with commitment.
    % as shown in Fig. \ref{fig:baseline2}. 
    Parameters are set as $s=0.1$; (a-b): $\sigma=0.5$; (c): $\varepsilon=0.1$ and $\sigma=0.5$; (d): $u=0.5$ and $\varepsilon=0.5$.\\}
    \label{figA3}
\end{figure}

\begin{figure}[htb]
\centering
\includegraphics[width=\linewidth]{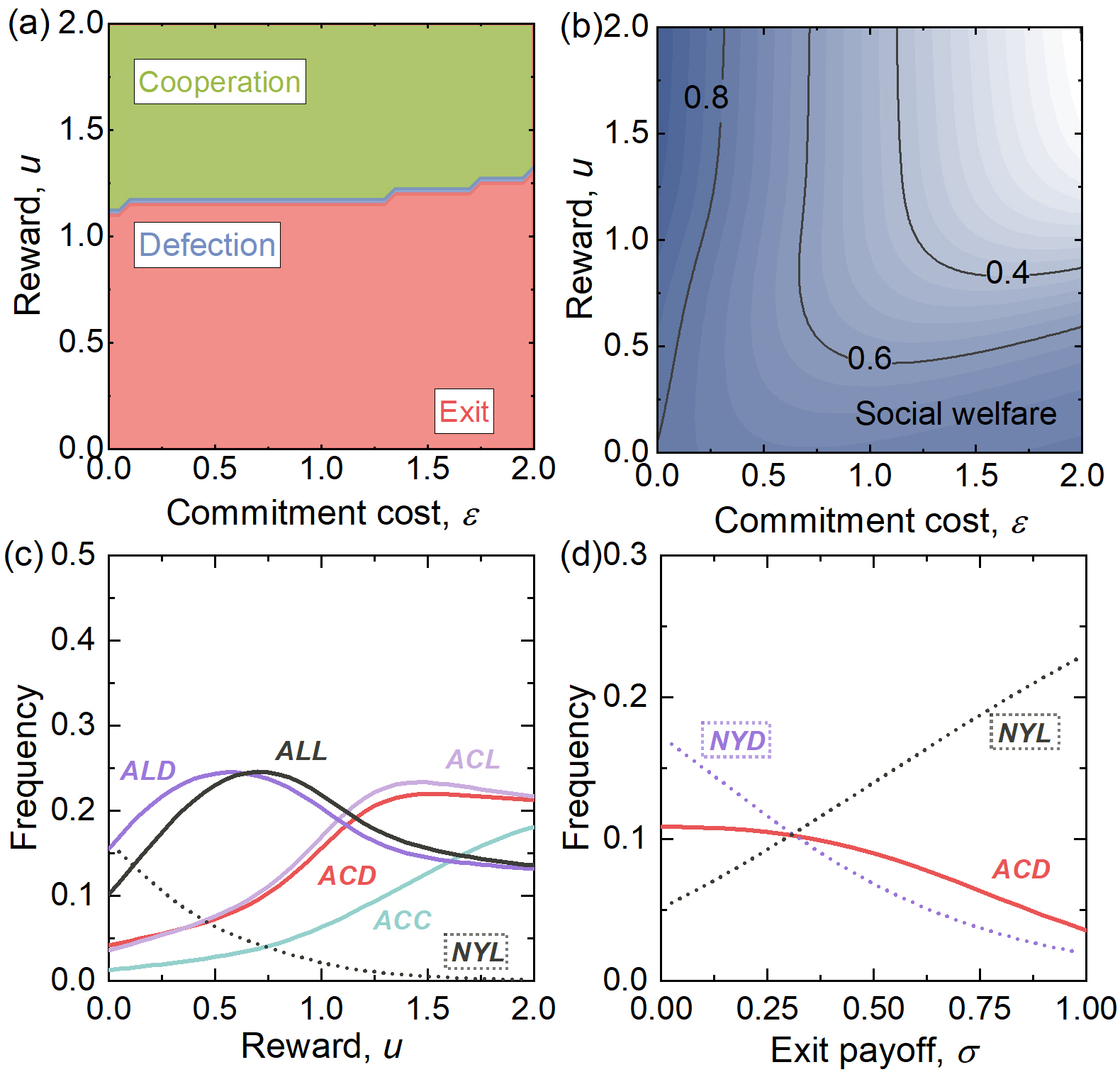}
    \caption{
    \textbf{
    FLEXIBLE-COM mechanism.}
    Panel (a) shows the region where cooperation, defection, and exit dominate in the population, respectively. 
    Panel (b) shows the social welfare as a function of commitment cost $\varepsilon$ and reward $u$.
    Panels (c) and (d) show the frequency of prevalent strategies (those with a frequency $>0.1$) as a function of reward and exit payoff, respectively.
    % The dark lines represent the results of OPD with commitment, while the light lines represent the results of the PD with commitment.
    % as shown in Fig. \ref{fig:baseline2}. 
    Parameters are set as $s=0.1$; (a-b): $\sigma=0.9$; (c): $\varepsilon=0.1$ and $\sigma=0.9$; (d): $u=0.5$ and $\varepsilon=0.9$.\\}
    \label{figA5}
\end{figure}

\begin{figure*}[tb]
\centering
\includegraphics[width=\linewidth]{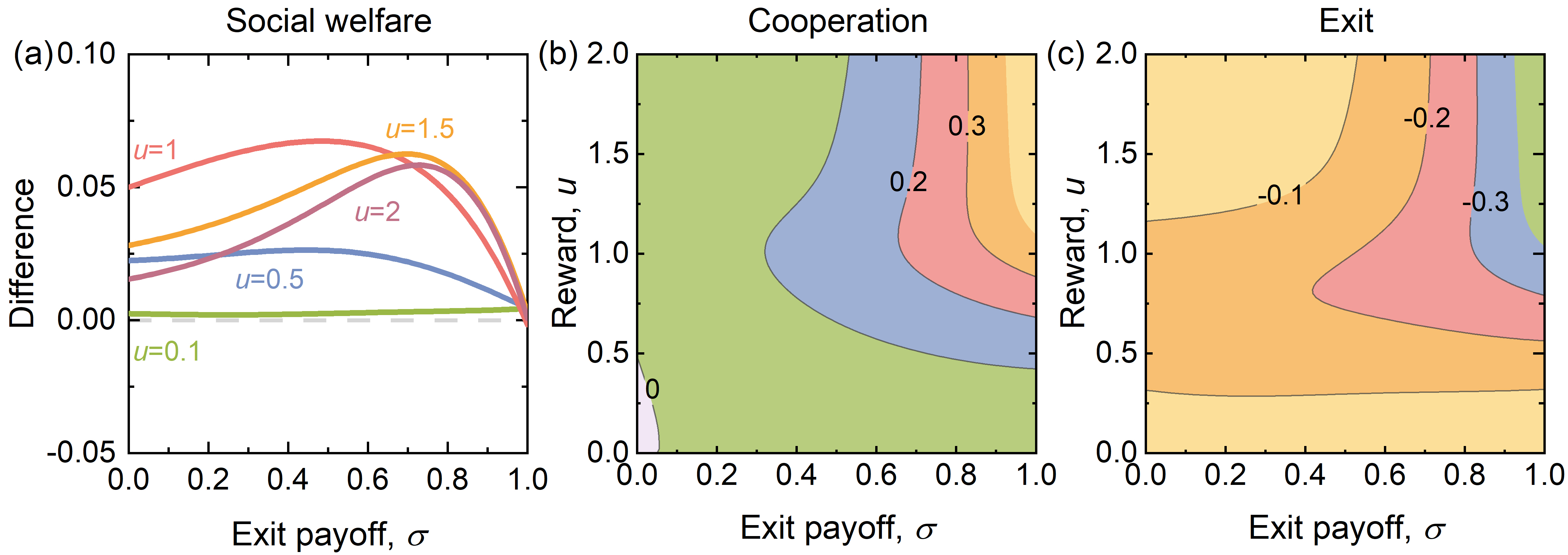}
    \caption{
    \textbf{Comparison between STRICT-COM and FLEXIBLE-COM.}
    Panel (a) shows the difference in social welfare between STRICT-COM and FLEXIBLE-COM as a function of exit payoff. Panels (b) and (c) show the difference in cooperation and exit between STRICT-COM and FLEXIBLE-COM as a function of exit payoff and reward, respectively. 
    Parameters are set as $\varepsilon=0.2$ and $s=0.1$.}
    % \vspace{0.2cm}
    \label{fig4-3}
\end{figure*}

% \newpage
% \section*{Appendix A}
% \setcounter{table}{0}
% \renewcommand{\thetable}{A\arabic{table}}

% \newpage 
\renewcommand{\arraystretch}{1.2}
\begin{sidewaystable}[htb]
  \centering
  \caption{Payoff Matrix without incentives}
  \label{PayoffMatrix}

  % Use tabularx to make the table automatically fill the page width.
  % The column types 'l' and 'C' are defined in the preamble of your document.
  \begin{tabularx}{\textheight}{l *{18}{C}}
    \toprule
    \textbf{} & \textbf{$ACC$} & \textbf{$ACD$} & \textbf{$ACL$} & \textbf{$ADC$} & \textbf{$ADD$} & \textbf{$ADL$} & \textbf{$ALC$} & \textbf{$ALD$} & \textbf{$ALL$} & \textbf{$NCC$} & \textbf{$NCD$} & \textbf{$NCL$} & \textbf{$NDC$} & \textbf{$NDD$} & \textbf{$NDL$} & \textbf{$NLC$} & \textbf{$NLD$} & \textbf{$NLL$} \\
    \midrule
    $ACC$ & $R'$ & $R'$ & $R'$ & $S'$ & $S'$ & $S'$ & $\sigma'$ & $\sigma'$ & $\sigma'$ & $R$ & $S$ & $\sigma$ & $R$ & $S$ & $\sigma$ & $R$ & $S$ & $\sigma$ \\
    $ACD$ & $R'$ & $R'$ & $R'$ & $S'$ & $S'$ & $S'$ & $\sigma'$ & $\sigma'$ & $\sigma'$ & $T$ & $P$ & $\sigma$ & $T$ & $P$ & $\sigma$ & $T$ & $P$ & $\sigma$ \\
    $ACL$ & $R'$ & $R'$ & $R'$ & $S'$ & $S'$ & $S'$ & $\sigma'$ & $\sigma'$ & $\sigma'$ & $\sigma$ & $\sigma$ & $\sigma$ & $\sigma$ & $\sigma$ & $\sigma$ & $\sigma$ & $\sigma$ & $\sigma$ \\
    \midrule
    $ADC$ & $T'$ & $T'$ & $T'$ & $P'$ & $P'$ & $P'$ & $\sigma'$ & $\sigma'$ & $\sigma'$ & $R$ & $S$ & $\sigma$ & $R$ & $S$ & $\sigma$ & $R$ & $S$ & $\sigma$ \\
    $ADD$ & $T'$ & $T'$ & $T'$ & $P'$ & $P'$ & $P'$ & $\sigma'$ & $\sigma'$ & $\sigma'$ & $T$ & $P$ & $\sigma$ & $T$ & $P$ & $\sigma$ & $T$ & $P$ & $\sigma$ \\
    $ADL$ & $T'$ & $T'$ & $T'$ & $P'$ & $P'$ & $P'$ & $\sigma'$ & $\sigma'$ & $\sigma'$ & $\sigma$ & $\sigma$ & $\sigma$ & $\sigma$ & $\sigma$ & $\sigma$ & $\sigma$ & $\sigma$ & $\sigma$ \\
    \midrule
    $ALC$ & $\sigma'$ & $\sigma'$ & $\sigma'$ & $\sigma'$ & $\sigma'$ & $\sigma'$ & $\sigma'$ & $\sigma'$ & $\sigma'$ & $R$ & $S$ & $\sigma$ & $R$ & $S$ & $\sigma$ & $R$ & $S$ & $\sigma$ \\
    $ALD$ & $\sigma'$ & $\sigma'$ & $\sigma'$ & $\sigma'$ & $\sigma'$ & $\sigma'$ & $\sigma'$ & $\sigma'$ & $\sigma'$ & $T$ & $P$ & $\sigma$ & $T$ & $P$ & $\sigma$ & $T$ & $P$ & $\sigma$ \\
    $ALL$ & $\sigma'$ & $\sigma'$ & $\sigma'$ & $\sigma'$ & $\sigma'$ & $\sigma'$ & $\sigma'$ & $\sigma'$ & $\sigma'$ & $\sigma$ & $\sigma$ & $\sigma$ & $\sigma$ & $\sigma$ & $\sigma$ & $\sigma$ & $\sigma$ & $\sigma$ \\
    \midrule
    $NCC$ & $R$ & $S$ & $\sigma$ & $R$ & $S$ & $\sigma$ & $R$ & $S$ & $\sigma$ & $R$ & $S$ & $\sigma$ & $R$ & $S$ & $\sigma$ & $R$ & $S$ & $\sigma$ \\
    $NCD$ & $T$ & $P$ & $\sigma$ & $T$ & $P$ & $\sigma$ & $T$ & $P$ & $\sigma$ & $T$ & $P$ & $\sigma$ & $T$ & $P$ & $\sigma$ & $T$ & $P$ & $\sigma$ \\
    $NCL$ & $\sigma$ & $\sigma$ & $\sigma$ & $\sigma$ & $\sigma$ & $\sigma$ & $\sigma$ & $\sigma$ & $\sigma$ & $\sigma$ & $\sigma$ & $\sigma$ & $\sigma$ & $\sigma$ & $\sigma$ & $\sigma$ & $\sigma$ & $\sigma$ \\
    \midrule
    $NDC$ & $R$ & $S$ & $\sigma$ & $R$ & $S$ & $\sigma$ & $R$ & $S$ & $\sigma$ & $R$ & $S$ & $\sigma$ & $R$ & $S$ & $\sigma$ & $R$ & $S$ & $\sigma$ \\
    $NDD$ & $T$ & $P$ & $\sigma$ & $T$ & $P$ & $\sigma$ & $T$ & $P$ & $\sigma$ & $T$ & $P$ & $\sigma$ & $T$ & $P$ & $\sigma$ & $T$ & $P$ & $\sigma$ \\
    $NDL$ & $\sigma$ & $\sigma$ & $\sigma$ & $\sigma$ & $\sigma$ & $\sigma$ & $\sigma$ & $\sigma$ & $\sigma$ & $\sigma$ & $\sigma$ & $\sigma$ & $\sigma$ & $\sigma$ & $\sigma$ & $\sigma$ & $\sigma$ & $\sigma$ \\
    \midrule
    $NLC$ & $R$ & $S$ & $\sigma$ & $R$ & $S$ & $\sigma$ & $R$ & $S$ & $\sigma$ & $R$ & $S$ & $\sigma$ & $R$ & $S$ & $\sigma$ & $R$ & $S$ & $\sigma$ \\
    $NLD$ & $T$ & $P$ & $\sigma$ & $T$ & $P$ & $\sigma$ & $T$ & $P$ & $\sigma$ & $T$ & $P$ & $\sigma$ & $T$ & $P$ & $\sigma$ & $T$ & $P$ & $\sigma$ \\
    $NLL$ & $\sigma$ & $\sigma$ & $\sigma$ & $\sigma$ & $\sigma$ & $\sigma$ & $\sigma$ & $\sigma$ & $\sigma$ & $\sigma$ & $\sigma$ & $\sigma$ & $\sigma$ & $\sigma$ & $\sigma$ & $\sigma$ & $\sigma$ & $\sigma$ \\
    \bottomrule
  \end{tabularx}
  \vspace{0.5cm}
  \footnotesize
  $R'=R-\varepsilon$, $S'=S-\varepsilon$, $T'=T-\varepsilon$, $P'=P-\varepsilon$, and $\sigma'=\sigma-\varepsilon$.
\end{sidewaystable}

\begin{sidewaystable}[htb]
  \centering
  \caption{Payoff Matrix with incentives for those who commit to cooperate before the game and cooperate in the game.}
  \label{table:incentive1}
  \setlength{\tabcolsep}{4pt} 
  \small 
  \begin{tabular}{l *{18}{c}}
    \toprule
    \textbf{} & \textbf{$ACC$} & \textbf{$ACD$} & \textbf{$ACL$} & \textbf{$ADC$} & \textbf{$ADD$} & \textbf{$ADL$} & \textbf{$ALC$} & \textbf{$ALD$} & \textbf{$ALL$} & \textbf{$NCC$} & \textbf{$NCD$} & \textbf{$NCL$} & \textbf{$NDC$} & \textbf{$NDD$} & \textbf{$NDL$} & \textbf{$NLC$} & \textbf{$NLD$} & \textbf{$NLL$} \\
    \midrule
    $ACC$ & $R'\!+\!u$ & $R'\!+\!u$ & $R'\!+\!u$ & $S'\!+\!u$ & $S'\!+\!u$ & $S'\!+\!u$ & $\sigma'\!+\!u$ & $\sigma'\!+\!u$ & $\sigma'\!+\!u$ & $R$ & $S$ & $\sigma$ & $R$ & $S$ & $\sigma$ & $R$ & $S$ & $\sigma$ \\
    $ACD$ & $R'\!+\!u$ & $R'\!+\!u$ & $R'\!+\!u$ & $S'\!+\!u$ & $S'\!+\!u$ & $S'\!+\!u$ & $\sigma'\!+\!u$ & $\sigma'\!+\!u$ & $\sigma'\!+\!u$ & $T$ & $P$ & $\sigma$ & $T$ & $P$ & $\sigma$ & $T$ & $P$ & $\sigma$ \\
    $ACL$ & $R'\!+\!u$ & $R'\!+\!u$ & $R'\!+\!u$ & $S'\!+\!u$ & $S'\!+\!u$ & $S'\!+\!u$ & $\sigma'\!+\!u$ & $\sigma'\!+\!u$ & $\sigma'\!+\!u$ & $\sigma$ & $\sigma$ & $\sigma$ & $\sigma$ & $\sigma$ & $\sigma$ & $\sigma$ & $\sigma$ & $\sigma$ \\
    \midrule
    $ADC$ & $T'$ & $T'$ & $T'$ & $P'$ & $P'$ & $P'$ & $\sigma'$ & $\sigma'$ & $\sigma'$ & $R$ & $S$ & $\sigma$ & $R$ & $S$ & $\sigma$ & $R$ & $S$ & $\sigma$ \\
    $ADD$ & $T'$ & $T'$ & $T'$ & $P'$ & $P'$ & $P'$ & $\sigma'$ & $\sigma'$ & $\sigma'$ & $T$ & $P$ & $\sigma$ & $T$ & $P$ & $\sigma$ & $T$ & $P$ & $\sigma$ \\
    $ADL$ & $T'$ & $T'$ & $T'$ & $P'$ & $P'$ & $P'$ & $\sigma'$ & $\sigma'$ & $\sigma'$ & $\sigma$ & $\sigma$ & $\sigma$ & $\sigma$ & $\sigma$ & $\sigma$ & $\sigma$ & $\sigma$ & $\sigma$ \\
    \midrule
    $ALC$ & $\sigma'$ & $\sigma'$ & $\sigma'$ & $\sigma'$ & $\sigma'$ & $\sigma'$ & $\sigma'$ & $\sigma'$ & $\sigma'$ & $R$ & $S$ & $\sigma$ & $R$ & $S$ & $\sigma$ & $R$ & $S$ & $\sigma$ \\
    $ALD$ & $\sigma'$ & $\sigma'$ & $\sigma'$ & $\sigma'$ & $\sigma'$ & $\sigma'$ & $\sigma'$ & $\sigma'$ & $\sigma'$ & $T$ & $P$ & $\sigma$ & $T$ & $P$ & $\sigma$ & $T$ & $P$ & $\sigma$ \\
    $ALL$ & $\sigma'$ & $\sigma'$ & $\sigma'$ & $\sigma'$ & $\sigma'$ & $\sigma'$ & $\sigma'$ & $\sigma'$ & $\sigma'$ & $\sigma$ & $\sigma$ & $\sigma$ & $\sigma$ & $\sigma$ & $\sigma$ & $\sigma$ & $\sigma$ & $\sigma$ \\
    \midrule
    $NCC$ & $R$ & $S$ & $\sigma$ & $R$ & $S$ & $\sigma$ & $R$ & $S$ & $\sigma$ & $R$ & $S$ & $\sigma$ & $R$ & $S$ & $\sigma$ & $R$ & $S$ & $\sigma$ \\
    $NCD$ & $T$ & $P$ & $\sigma$ & $T$ & $P$ & $\sigma$ & $T$ & $P$ & $\sigma$ & $T$ & $P$ & $\sigma$ & $T$ & $P$ & $\sigma$ & $T$ & $P$ & $\sigma$ \\
    $NCL$ & $\sigma$ & $\sigma$ & $\sigma$ & $\sigma$ & $\sigma$ & $\sigma$ & $\sigma$ & $\sigma$ & $\sigma$ & $\sigma$ & $\sigma$ & $\sigma$ & $\sigma$ & $\sigma$ & $\sigma$ & $\sigma$ & $\sigma$ & $\sigma$ \\
    \midrule
    $NDC$ & $R$ & $S$ & $\sigma$ & $R$ & $S$ & $\sigma$ & $R$ & $S$ & $\sigma$ & $R$ & $S$ & $\sigma$ & $R$ & $S$ & $\sigma$ & $R$ & $S$ & $\sigma$ \\
    $NDD$ & $T$ & $P$ & $\sigma$ & $T$ & $P$ & $\sigma$ & $T$ & $P$ & $\sigma$ & $T$ & $P$ & $\sigma$ & $T$ & $P$ & $\sigma$ & $T$ & $P$ & $\sigma$ \\
    $NDL$ & $\sigma$ & $\sigma$ & $\sigma$ & $\sigma$ & $\sigma$ & $\sigma$ & $\sigma$ & $\sigma$ & $\sigma$ & $\sigma$ & $\sigma$ & $\sigma$ & $\sigma$ & $\sigma$ & $\sigma$ & $\sigma$ & $\sigma$ & $\sigma$ \\
    \midrule
    $NLC$ & $R$ & $S$ & $\sigma$ & $R$ & $S$ & $\sigma$ & $R$ & $S$ & $\sigma$ & $R$ & $S$ & $\sigma$ & $R$ & $S$ & $\sigma$ & $R$ & $S$ & $\sigma$ \\
    $NLD$ & $T$ & $P$ & $\sigma$ & $T$ & $P$ & $\sigma$ & $T$ & $P$ & $\sigma$ & $T$ & $P$ & $\sigma$ & $T$ & $P$ & $\sigma$ & $T$ & $P$ & $\sigma$ \\
    $NLL$ & $\sigma$ & $\sigma$ & $\sigma$ & $\sigma$ & $\sigma$ & $\sigma$ & $\sigma$ & $\sigma$ & $\sigma$ & $\sigma$ & $\sigma$ & $\sigma$ & $\sigma$ & $\sigma$ & $\sigma$ & $\sigma$ & $\sigma$ & $\sigma$ \\
    \bottomrule
  \end{tabular}
  \vspace{0.5cm}
  \footnotesize
  $R'=R-\varepsilon$, $S'=S-\varepsilon$, $T'=T-\varepsilon$, $P'=P-\varepsilon$, and $\sigma'=\sigma-\varepsilon$.
\end{sidewaystable}

\begin{sidewaystable}[htb]
  \centering
  \caption{Payoff Matrix with incentives for those who commit not to defect before the game and cooperate or exit in the game.}
  \label{table:incentive2}
  \setlength{\tabcolsep}{4pt} 
  \small 
  \begin{tabular}{l *{18}{c}}
    \toprule
    \textbf{} & \textbf{$ACC$} & \textbf{$ACD$} & \textbf{$ACL$} & \textbf{$ADC$} & \textbf{$ADD$} & \textbf{$ADL$} & \textbf{$ALC$} & \textbf{$ALD$} & \textbf{$ALL$} & \textbf{$NCC$} & \textbf{$NCD$} & \textbf{$NCL$} & \textbf{$NDC$} & \textbf{$NDD$} & \textbf{$NDL$} & \textbf{$NLC$} & \textbf{$NLD$} & \textbf{$NLL$} \\
    \midrule
    $ACC$ & $R'\!\!\!+\!u$ & $R'\!\!\!+\!u$ & $R'\!\!\!+\!u$ & $S'\!\!\!+\!u$ & $S'\!\!\!+\!u$ & $S'\!\!\!+\!u$ & $\sigma'\!\!\!+\!u$ & $\sigma'\!\!\!+\!u$ & $\sigma'\!\!\!+\!u$ & $R$ & $S$ & $\sigma$ & $R$ & $S$ & $\sigma$ & $R$ & $S$ & $\sigma$ \\
    $ACD$ & $R'\!\!\!+\!u$ & $R'\!\!\!+\!u$ & $R'\!\!\!+\!u$ & $S'\!\!\!+\!u$ & $S'\!\!\!+\!u$ & $S'\!\!\!+\!u$ & $\sigma'\!\!\!+\!u$ & $\sigma'\!\!\!+\!u$ & $\sigma'\!\!\!+\!u$ & $T$ & $P$ & $\sigma$ & $T$ & $P$ & $\sigma$ & $T$ & $P$ & $\sigma$ \\
    $ACL$ & $R'\!\!\!+\!u$ & $R'\!\!\!+\!u$ & $R'\!\!\!+\!u$ & $S'\!\!\!+\!u$ & $S'\!\!\!+\!u$ & $S'\!\!\!+\!u$ & $\sigma'\!\!\!+\!u$ & $\sigma'\!\!\!+\!u$ & $\sigma'\!\!\!+\!u$ & $\sigma$ & $\sigma$ & $\sigma$ & $\sigma$ & $\sigma$ & $\sigma$ & $\sigma$ & $\sigma$ & $\sigma$ \\
    \midrule
    $ADC$ & $T'$ & $T'$ & $T'$ & $P'$ & $P'$ & $P'$ & $\sigma'$ & $\sigma'$ & $\sigma'$ & $R$ & $S$ & $\sigma$ & $R$ & $S$ & $\sigma$ & $R$ & $S$ & $\sigma$ \\
    $ADD$ & $T'$ & $T'$ & $T'$ & $P'$ & $P'$ & $P'$ & $\sigma'$ & $\sigma'$ & $\sigma'$ & $T$ & $P$ & $\sigma$ & $T$ & $P$ & $\sigma$ & $T$ & $P$ & $\sigma$ \\
    $ADL$ & $T'$ & $T'$ & $T'$ & $P'$ & $P'$ & $P'$ & $\sigma'$ & $\sigma'$ & $\sigma'$ & $\sigma$ & $\sigma$ & $\sigma$ & $\sigma$ & $\sigma$ & $\sigma$ & $\sigma$ & $\sigma$ & $\sigma$ \\
    \midrule
    $ALC$ & $\sigma'\!\!+\!u$ & $\sigma'\!\!+\!u$ & $\sigma'\!\!+\!u$ & $\sigma'\!\!+\!u$ & $\sigma'\!\!+\!u$ & $\sigma'\!\!+\!u$ & $\sigma'\!\!+\!u$ & $\sigma'\!\!+\!u$ & $\sigma'\!\!+\!u$ & $R$ & $S$ & $\sigma$ & $R$ & $S$ & $\sigma$ & $R$ & $S$ & $\sigma$ \\
    $ALD$ & $\sigma'\!\!+\!u$ & $\sigma'\!\!+\!u$ & $\sigma'\!\!+\!u$ & $\sigma'\!\!+\!u$ & $\sigma'\!\!+\!u$ & $\sigma'\!\!+\!u$ & $\sigma'\!\!+\!u$ & $\sigma'\!\!+\!u$ & $\sigma'\!\!+\!u$  & $T$ & $P$ & $\sigma$ & $T$ & $P$ & $\sigma$ & $T$ & $P$ & $\sigma$ \\
    $ALL$ & $\sigma'\!\!+\!u$ & $\sigma'\!\!+\!u$ & $\sigma'\!\!+\!u$ & $\sigma'\!\!+\!u$ & $\sigma'\!\!+\!u$ & $\sigma'\!\!+\!u$ & $\sigma'\!\!+\!u$ & $\sigma'\!\!+\!u$ & $\sigma'\!\!+\!u$  & $\sigma$ & $\sigma$ & $\sigma$ & $\sigma$ & $\sigma$ & $\sigma$ & $\sigma$ & $\sigma$ & $\sigma$ \\
    \midrule
    $NCC$ & $R$ & $S$ & $\sigma$ & $R$ & $S$ & $\sigma$ & $R$ & $S$ & $\sigma$ & $R$ & $S$ & $\sigma$ & $R$ & $S$ & $\sigma$ & $R$ & $S$ & $\sigma$ \\
    $NCD$ & $T$ & $P$ & $\sigma$ & $T$ & $P$ & $\sigma$ & $T$ & $P$ & $\sigma$ & $T$ & $P$ & $\sigma$ & $T$ & $P$ & $\sigma$ & $T$ & $P$ & $\sigma$ \\
    $NCL$ & $\sigma$ & $\sigma$ & $\sigma$ & $\sigma$ & $\sigma$ & $\sigma$ & $\sigma$ & $\sigma$ & $\sigma$  & $\sigma$ & $\sigma$ & $\sigma$ & $\sigma$ & $\sigma$ & $\sigma$ & $\sigma$ & $\sigma$ & $\sigma$ \\
    \midrule
    $NDC$ & $R$ & $S$ & $\sigma$ & $R$ & $S$ & $\sigma$ & $R$ & $S$ & $\sigma$ & $R$ & $S$ & $\sigma$ & $R$ & $S$ & $\sigma$ & $R$ & $S$ & $\sigma$ \\
    $NDD$ & $T$ & $P$ & $\sigma$ & $T$ & $P$ & $\sigma$ & $T$ & $P$ & $\sigma$ & $T$ & $P$ & $\sigma$ & $T$ & $P$ & $\sigma$ & $T$ & $P$ & $\sigma$ \\
    $NDL$ & $\sigma$ & $\sigma$ & $\sigma$ & $\sigma$ & $\sigma$ & $\sigma$ & $\sigma$ & $\sigma$ & $\sigma$  & $\sigma$ & $\sigma$ & $\sigma$ & $\sigma$ & $\sigma$ & $\sigma$ & $\sigma$ & $\sigma$ & $\sigma$ \\
    \midrule
    $NLC$ & $R$ & $S$ & $\sigma$ & $R$ & $S$ & $\sigma$ & $R$ & $S$ & $\sigma$ & $R$ & $S$ & $\sigma$ & $R$ & $S$ & $\sigma$ & $R$ & $S$ & $\sigma$ \\
    $NLD$ & $T$ & $P$ & $\sigma$ & $T$ & $P$ & $\sigma$ & $T$ & $P$ & $\sigma$ & $T$ & $P$ & $\sigma$ & $T$ & $P$ & $\sigma$ & $T$ & $P$ & $\sigma$ \\
    $NLL$ & $\sigma$ & $\sigma$ & $\sigma$ & $\sigma$ & $\sigma$ & $\sigma$ & $\sigma$ & $\sigma$ & $\sigma$  & $\sigma$ & $\sigma$ & $\sigma$ & $\sigma$ & $\sigma$ & $\sigma$ & $\sigma$ & $\sigma$ & $\sigma$ \\
    \bottomrule
  \end{tabular}
  \vspace{0.5cm}
  \footnotesize
  $R'=R-\varepsilon$, $S'=S-\varepsilon$, $T'=T-\varepsilon$, $P'=P-\varepsilon$, and $\sigma'=\sigma-\varepsilon$.
\end{sidewaystable}

% \begin{figure*}[tb]
% \centering
% \includegraphics[width=\linewidth]{fig4-1.jpg}
%     \caption{
%     \textbf{Comparison between STRICT-COM and FLEXIBLE-COM.}
%     Panel (a) shows the difference in social welfare between STRICT-COM and FLEXIBLE-COM as a function of exit payoff. Panels (b) and (c) show the difference in cooperation and exit between STRICT-COM and FLEXIBLE-COM as a function of exit payoff and reward, respectively. 
%     Parameters are set as $\varepsilon=0.5$ and $s=0.1$.}
%     % \vspace{0.2cm}
%     \label{fig4-1}
% \end{figure*}
% \begin{figure*}[tb]
% \centering
% \includegraphics[width=\linewidth]{fig4-2.jpg}
%     \caption{
%     \textbf{Comparison between STRICT-COM and FLEXIBLE-COM.}
%     Panel (a) shows the difference in social welfare between STRICT-COM and FLEXIBLE-COM as a function of exit payoff. Panels (b) and (c) show the difference in cooperation and exit between STRICT-COM and FLEXIBLE-COM as a function of exit payoff and reward, respectively. 
%     Parameters are set as $\varepsilon=0.9$ and $s=0.1$.}
%     % \vspace{0.2cm}
%     \label{fig4-2}
% \end{figure*}

\end{document}